\documentclass[12pt]{article}
\usepackage{aaspp4}
\usepackage{amstex}
\usepackage{natbib}
\usepackage{psfig}
\def\eg{{\it e.g.}\ }
\def\ie{{\it i.e.}\ }

\def\lae{\mathrel{<\kern-1.0em\lower0.9ex\hbox{$\sim$}}}

\def\gae{\mathrel{>\kern-1.0em\lower0.9ex\hbox{$\sim$}}}

\newcommand{\abox}{\mathrel{\sqcup\kern-0.70em\hbox{$\sqcap$}}}

\def\25{25~$\mu$m}
\def\60{60~$\mu$m}
\def\100{100~$\mu$m}

\newdimen\rotdimen
\def\vspec#1{\special{ps:#1}}%  passes #1 verbatim to the output
\def\rotstart#1{\vspec{gsave currentpoint currentpoint translate
   #1 neg exch neg exch translate}}% #1 can be any origin-fixing transformation
\def\rotfinish{\vspec{currentpoint grestore moveto}}% gets back in synch
\def\rotr#1{\rotdimen=\ht#1\advance\rotdimen by\dp#1%
   \hbox to\rotdimen{\hskip\ht#1\vbox to\wd#1{\rotstart{90 rotate}%
   \box#1\vss}\hss}\rotfinish}
\def\rotl#1{\rotdimen=\ht#1\advance\rotdimen by\dp#1%
   \hbox to\rotdimen{\vbox to\wd#1{\vskip\wd#1\rotstart{270 rotate}%
   \box#1\vss}\hss}\rotfinish}%
\def\rotu#1{\rotdimen=\ht#1\advance\rotdimen by\dp#1%
   \hbox to\wd#1{\hskip\wd#1\vbox to\rotdimen{\vskip\rotdimen
   \rotstart{-1 dup scale}\box#1\vss}\hss}\rotfinish}%
\def\rotf#1{\hbox to\wd#1{\hskip\wd#1\rotstart{-1 1 scale}%
   \box#1\hss}\rotfinish}%

\newcommand{\qv}{{\em q.v. }}
\newcommand{\cf}{{\em c.f. }}
\begin{document}

\title{Cosmological Parameters from Multiple-arc Gravitational Lensing 
Systems I: smooth lensing potentials}

\author{Robert Link\altaffilmark{1} \& Michael J. Pierce}
\affil{Indiana University, Department of Astronomy, Swain West Rm. 319,
Bloomington, IN 47405}

\altaffiltext{1}{In partial fulfillment of a Ph.D. Dissertation in Astronomy 
at Indiana University}

\newcommand{\ebar}{\bar e}
\newcommand{\ahat}{\hat\alpha}
\newcommand{\ahatb}{\hat{\boldsymbol{\alpha}}}

%\bibliographystyle{dcu}
%\citationstyle{agsm}

\begin{abstract}

We describe a new approach for the determination of cosmological
parameters using gravitational lensing systems with multiple arcs.  We
exploit the fact that a given cluster can produce multiple arcs from
sources over a broad range in redshift.  The coupling between the
critical radius of a single arc and the projected mass density of the
lensing cluster can be avoided by considering the relative positions
of two or more arcs. Cosmological sensitivity appears through the
angular size-redshift relation.  We present a simple analytic argument
for this approach using an axisymmetric, power-law cluster potential.
In this case, the relative positions of the arcs can be shown to
depend only upon the ratios of the angular-size distances of the
source and between the lens and source.  Provided that the astrometric
precision approaches $\sim$ 0.01 arcsec (\eg via HST) and the
redshifts of the arcs are known, we show that the system can, in
principle, provide cosmological information through the angular-size
redshift relation.

We next consider simulated data constructed using a more general form
for the potential, realistic sources, and an assumed cosmology.  We
present a method for simultaneously inverting the lens and extracting
the cosmological parameters. The input data required are the image and
measured redshifts for the arcs. The technique relies upon the
conservation of surface brightness in gravitationally lensed systems.
We find that for a simple lens model our approach can recover the
cosmological parameters assumed in the construction of the simulated
images.

\end{abstract}


\section{Introduction}

One particularly promising approach for measuring the large-scale
geometry of the universe is through the exploitation of gravitational
lensing of distant galaxies and quasars.  For some time gravitational
lensing has been examined as a possible means of constraining
cosmological parameters such as $q_0$ and $\lambda$.
%\cite[]{Ref64,Pac81,Tur90}.  
\cite[\eg][]{Tur90} To date most methods have focused on statistical
arguments based on the cross sections for strong lensing of background
objects \cite[\eg][]{Cen94}.  As an alternative we are investigating
the feasibility of a method which makes more direct use of the
dependence of gravitational lensing events on the geometry of the
spacetime between the observer, the source and the lens.  The method
is motivated by recent images of cluster systems containing numerous
arcs \cite[\eg][]{Lup94,Kne96}.  The method works by requiring
consistency between both a single mass model for the cluster and a
single angular-size redshift relation for all the lensed sources.

We begin with an overview of gravitational lensing, followed by an
analytic argument assuming an axisymmetric potential for the lensing
cluster given by a power-law in density.  We show that the astrometric
precision afforded by HST is sufficient to distinguish between
differing cosmologies through the angular-size redshift relation.  We
then investigate more realistic models for the lensing cluster and
sources.  However, in these cases the resulting arcs have more
complicated morphology such that the lens and the geometry must be
inverted via ray tracing and the minimization of a suitable merit
function (\eg one based upon the conservation of surface brightness
and reasonable morphology of the sources, etc.) \cite[]{Koc89}.  
%xxxChange 6
In section~\ref{sec:method} we present an analysis of simulated data
generated with an assumed form for the lensing potential and an
arbitrary cosmology.  With the simulated image and the redshifts of
the arcs as input data we find that we can recover both the parameters
describing the lensing mass distribution and the cosmology.  In
section~\ref{sec:rslts} we discuss the results of this analysis.  We
develop estimates of the bounds that could be set on cosmological
parameters from a hypothetical system with the characteristics of our
simulation, and we argue that this technique is a good complement to
existing methods for the determination of cosmological parameters
because it is purely geometrical rather than statistical in nature and
is insensitive to evolutionary effects.  In the concluding section we
summarize the results of the previous sections, and we describe
further refinements to our analysis which will be detailed in
subsequent papers in this series.

\section{Gravitational Lensing of Multiple Sources}

\begin{figure}
\hbox{%
\psfig{figure=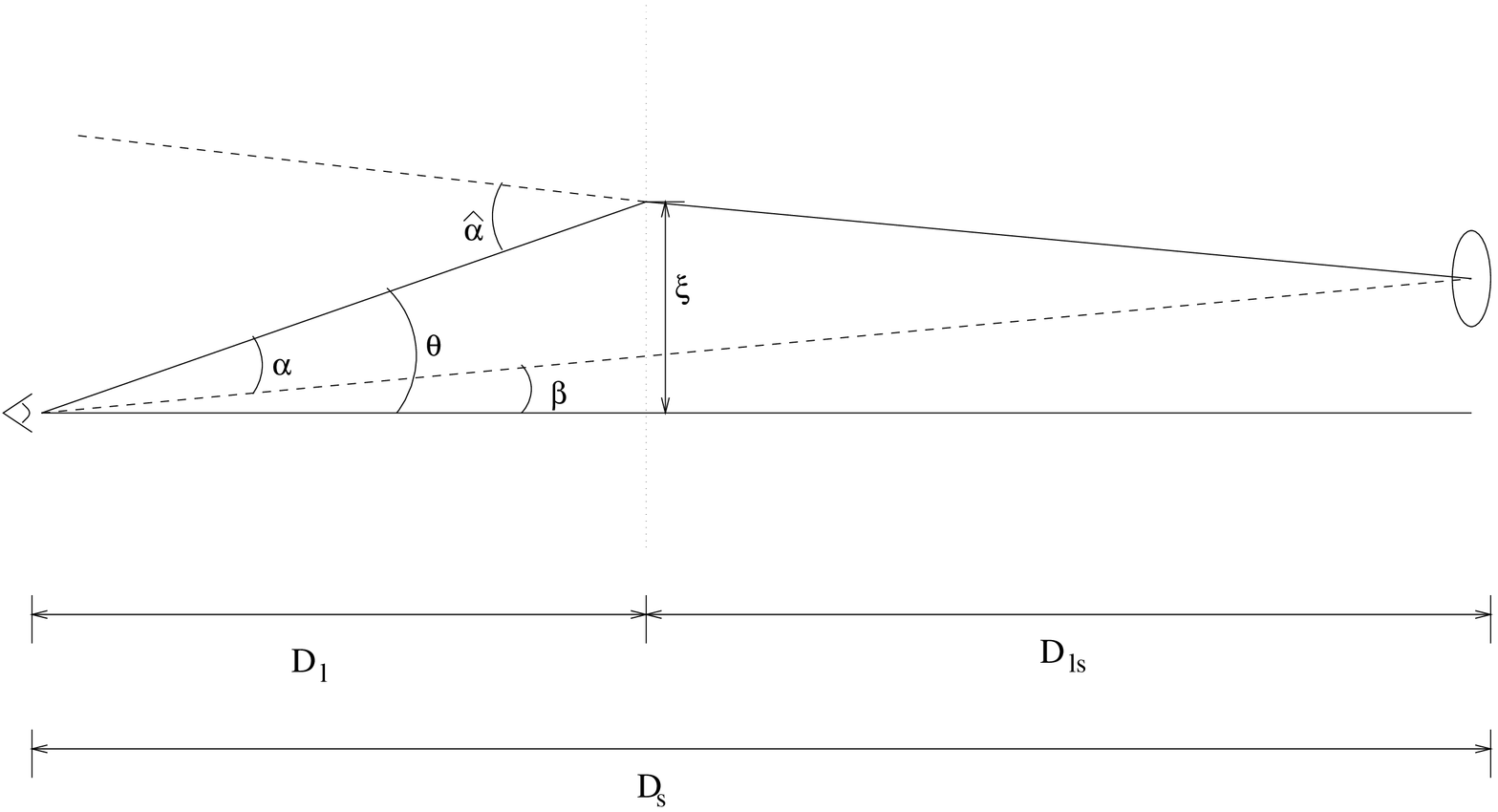,width=5in,angle=-90}}
\caption{The geometry of gravitational lensing in the thin lens
approximation.  For a light ray passing through the lens plane at
impact parameter $\xi$, the reduced deflection angle ($\alpha$) is
given by the gradient of the lensing potential.  The undeflected
position of the source ($\beta$) is related to the observed position
($\theta$) by the lens mapping (equation~\protect\ref{eq:lensmap} in
text).  All distances are angular size distances.}
\label{fig:lensgeom}
\end{figure}

Gravitational lensing is sensitive to the gravitational potential of
the lens, the angular position of the source object relative to the
lens, and the angular size distances between observer, lens, and
source (fig.~\ref{fig:lensgeom}).  This last is of principal interest,
since the dependence of angular size distance on redshift has a direct
dependence on cosmological parameters \cite[\eg][]{San88}.  Large
arcs are produced by the near alignment of distant sources (\ie
galaxies) and an intervening cluster of galaxies
\cite[]{Pac87,Sou87,Lyn88}.  For an axisymmetric (thin) lens with an
on-axis source, the ``critical radius'' at which the source is
observed depends only upon the projected surface mass density and the
angular size distances to the lens and the source \cite[]{Bla92}.
Thus, by measuring redshifts for the gravitationally lensed sources
and explicitly including the cosmological dependence of angular size
distance on those redshifts, constraints could be set upon various
cosmological models.  We believe that the recently observed cluster
systems with several multiply imaged arcs offer one of the best
opportunities for obtaining meaningful constraints.  We exploit the
fact that sources distributed over a range in redshift can be imaged
by the same lens.  This should allow us to separate the effects of the
lens mass distribution from the cosmological effects.

The technique we propose involves inverting the lens system using
methods similar to those presented in \cite{Koc89}, but with one
important difference; multiple sources at different redshifts are
included.  Inclusion of multiple source planes admits the possibility
of cosmological dependence (\qv
\S\S~\ref{sec:analytic},\ref{sec:potential}), which is treated in
our models by including the cosmological parameters as additional
fitting parameters.  For each cosmological model being tested the
distances to all of the sources are computed using the redshifts and
the particular cosmological model under consideration.  The distances
so calculated affect the quality of the fit, and this gives our method
its cosmological sensitivity.  While this does increase the number of
free parameters in our model, we will show that the existence of
multiple multiply imaged sources allows the modeled distances to be
recovered with sufficient accuracy to provide interesting constraints
on cosmological parameters.

\subsection{Analytic Treatment of an Axisymmetric System}
\label{sec:analytic}
%xxx Change 1
We can gain an understanding of the cosmological sensitivity of
multiple-arc gravitational lens systems by considering a simple, if
somewhat contrived, analytically treatable model.  Accordingly,
consider a cluster potential approximated as an axisymmetric power law
in density.
\begin{equation}
\label{eq:isosphere}
\rho(r) = \rho_a \left(\frac{a}{r}\right)^b.
\end{equation} 
The projected mass density is then
\begin{equation}
\label{eq:projmass}
\Sigma_M(R) = \rho_a a^b {\cal I} R^{1-b},
\end{equation}
where 
\begin{equation}
{\cal I} \equiv \int_{-\frac{\pi}{2}}^{\frac{\pi}{2}} d\theta
cos^{b-2} \theta,
\end{equation}
which is independent of $R$.
Integrating equation~(\ref{eq:projmass}) gives the total projected mass
within the projected radius, $R$.
\begin{equation}
\label{eq:projmass2}
M(R) = 2\pi\rho_a a^b \frac{{\cal I}}{3-b} R^{3-b}.
\end{equation}
For $b=2$ this equation reduces to the familiar isothermal case
\cite[\cf][]{Bre92}.

%xxx Change 2
For simplicity, consider the case of on-axis sources, which would
appear as ``Einstein Rings'' at the angular critical radii,
$\theta_E$.
Then
\begin{equation}
\label{eq:einsteinring}
\theta_E^2 = \frac{4G_N M(\theta_E)}{c^2} \frac{D_{ls}}{D_l D_s},
\end{equation}
where $D_l$, $D_s$, and $D_{ls}$ are the angular size distances of the
lens, source, and lens-to-source, respectively \cite[\cf][ \protect\S
2.1]{Sch92}.  Since $R \equiv D_l \theta$ we can combine equations
(\ref{eq:projmass2}) and (\ref{eq:einsteinring}) such that:
\begin{equation}
\theta_E^2 = \frac{8\pi G_N}{c^2} \rho_a a^b \frac{{\cal I}}{3-b}
D_l^{2-b} \frac{D_{ls}}{D_s} \theta_E^{3-b},
\end{equation}
whence
\begin{equation}
\label{eq:thetae}
\theta_E = \left(\frac{8\pi G_N}{c^2}\rho_a a^b \frac{{\cal I}}{3-b}
D_l^{2-b}\right)^{\frac{1}{b-1}}
\left(\frac{D_{ls}}{D_s}\right)^{\frac{1}{b-1}}.
\end{equation}
Thus, the angular radius of a given arc depends upon both the
characteristics of the lens object and the angular-size distances to
the source and lens-to-source.  Note that for a particular lens system
with multiple sources at differing distances the first expression on
the left hand side of equation~(\ref{eq:thetae}) is fixed, and only
the second varies between sources.

Now consider the case of multiple arcs.  The ratio of the Einstein
radii for two sources, $k$ and $m$ is
\begin{equation}
\frac{\theta_k}{\theta_m} =
\left(\frac{D_{ls}^{(k)}}{D_s^{(k)}}\right)^{\frac{1}{b-1}}
\left(\frac{D_{s}^{(m)}}{D_{ls}^{(m)}}\right)^{\frac{1}{b-1}}.
\end{equation}
We can combine this relation for three arcs, $k,m, \text{and}\ n$ to
eliminate $b$.
\begin{equation}
\label{eq:predictions}
\frac{\ln\left(\frac{\theta_k}{\theta_m}\right)}%
{\ln\left(\frac{\theta_k}{\theta_n}\right)}
=
\frac{\ln\left(\frac{D_{ls}^{(k)}D_s^{(m)}}{D_s^{(k)}D_{ls}^{(m)}}\right)}%
{\ln\left(\frac{D_{ls}^{(k)}D_s^{(n)}}{D_s^{(k)}D_{ls}^{(n)}}\right)}.
\end{equation}
The left side of equation~(\ref{eq:predictions}) is an observational
quantity which can in principle be measured directly, and the right
side is a cosmological factor which can be calculated for any desired
cosmological model, given the redshifts of the arcs.  Thus, the
relative radii of giant arcs has cosmological sensitivity and can
provide a direct test of cosmological models.

Standard error propagation techniques can be applied to
equation~(\ref{eq:predictions}) to determine the astrometric precision
required to allow sensitivity to cosmological effects.  Let the
angular radii, $\theta$ of the arcs be measurable with errors of $\pm
\sigma$.  Then the error in
\begin{equation}
{\cal R} \equiv
\frac{\ln\left(\frac{\theta_k}{\theta_m}\right)}%
{\ln\left(\frac{\theta_k}{\theta_n}\right)}
\end{equation}
is given by
\begin{equation}
\label{eq:err}
\sigma_{\cal R}^2 = \frac{\sigma^2}{\left(\ln
\left(\frac{\theta_k}{\theta_n}\right)\right)^2} 
\left[\frac{1}{\theta_k^2} - \frac{1}{\theta_m^2} - 
\left(\frac{\ln\left(\frac{\theta_k}{\theta_m}\right)}%
{\ln\left(\frac{\theta_k}{\theta_n}\right)}\right)^2
\left(\frac{1}{\theta_k^2} - \frac{1}{\theta_n^2}\right)\right].
\end{equation}
We can evaluate equation~(\ref{eq:err}) for a typical lens system,
assuming a standard $\Omega=1$, $\lambda=0$ cosmology.
Consider a lens system ($z_l = 0.4$) with three arcs ($z_{1,2,3} =
0.6,0.8,1.2$).  For $\theta(z\rightarrow\infty) = 30\arcsec$ the
system produces arcs with angular radii $\theta_{1,2,3} =
7\farcs8,11\farcs8,15\farcs7$. 
These values in equation~(\ref{eq:err}) yield the result
\begin{equation}
\sigma_{\cal R} \approx 0.1 \sigma\text{(arcsec)}.
\end{equation}

\begin{figure}
\hbox{%
\psfig{figure=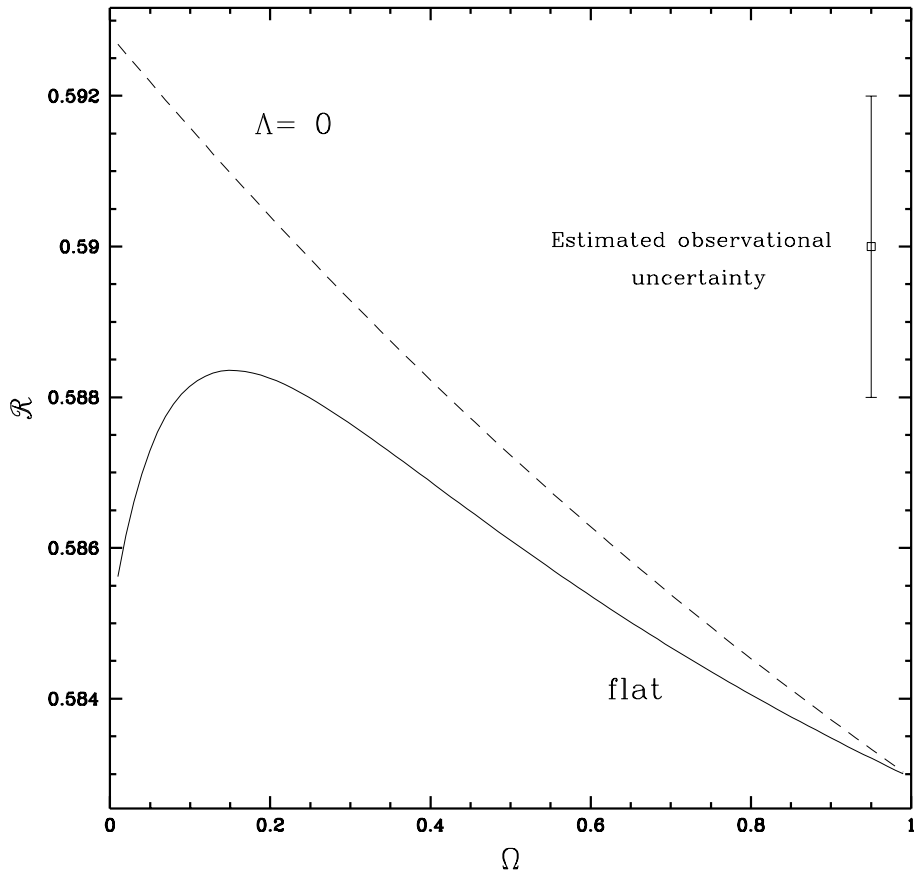,width=5in}}
\caption{The dependence of relative angular radii of gravitational
arcs on cosmological parameters for an analytically soluble
axisymmetric lens model (see \S~\protect\ref{sec:analytic}).  Here
${\cal R} =
\ln\left(\frac{\theta_1}{\theta_2}\right)/\ln\left(\frac{\theta_1}{\theta_3}\right)$.
Computed values of ${\cal R}$ are shown for two families of
cosmological models.  For comparison the we include a conservative
estimate of the uncertainty for measurements obtainable using HST.}
\label{fig:anly}
\end{figure}

The right hand side of equation~(\ref{eq:predictions}) can be
evaluated for any desired cosmology.  Figure~\ref{fig:anly} shows
calculated values of ${\cal R}$ for two families of cosmological
models: flat cosmologies and $\lambda = 0$ cosmologies.  From these we
can conclude that a measurement error of $\sigma_{\cal R} \approx
0.001$ would allow marginal discrimination between extreme
cosmological cases.  This corresponds to a measurement error in the
positions of the arcs of $\sigma_{\text{arc}} \approx 0\farcs 01$.

We can compare the results of this section to the astrometric
precision one might expect to obtain from presently available
instruments.  Examination of arcs imaged by HST shows that the cross
section of a representative arc has a FWHM of $0\farcs 2$.  The arcs
have a typical signal-to-noise ratio of 10.  A simple monte carlo
calculation indicates that the positions of the arcs should be
measurable to an error of $\sigma_{\text{arc}} \approx 0\farcs 02$.
Such a measurement would result in marginal discrimination between
cosmological models (fig.~\ref{fig:anly}).  

%xxx Change 3
Real gravitational lens systems are expected to be more complex than
these analytically soluble systems.  Moreover, even for the simple
potentials we have discussed, the off-axis sources present in real
systems introduce an additional error term which can dominate the
contribution to the total error.  Consequently, the technique of
comparing the astrometric positions of the observed arcs to their
calculated positions for the models being tested is ill-suited for
application to real systems.  In section~\ref{sec:method} we describe
a numerical technique which addresses both of these issues.

\section{Evaluation of More Realistic Cases Using Simulated Data}
\label{sec:method}

%xxx Change 3 (cont.)
The surface brightness of a bundle of light rays is unchanged by
gravitational deflection \cite[]{Bla92}.  We can exploit this property
to construct a technique for analyzing systems more realistic than
that described in section~\ref{sec:analytic}.  With this method
comparing the surface brightnesses of image pixels mapped to the same
point in the source plane replaces the measurement of arc positions as
our primary test of the quality of a model.

\subsection{Lensing Potential}
\label{sec:potential}

The \emph{lensing potential} is defined by
\begin{equation}
\label{eq:deflangl}
\boldsymbol{\alpha}(\boldsymbol{\theta}) = \boldsymbol{\nabla} 
\psi(\boldsymbol{\theta}),
\end{equation}
where $\boldsymbol{\alpha}$ is the reduced deflection angle depicted in
figure~\ref{fig:lensgeom}, and $\boldsymbol{\theta}$ is the position
angle at which the source is observed.
Thus, we see that the \emph{lens mapping} is defined by
\begin{equation}
\label{eq:lensmap}
\begin{split}
\boldsymbol{\beta} &= \boldsymbol{\theta} - \boldsymbol{\alpha} \\
                   &= \boldsymbol{\theta} - 
                      \boldsymbol{\nabla}\psi(\boldsymbol{\theta}).
\end{split}
\end{equation}

We adopt an elliptical lensing potential of the form \cite[\eg][]{Bla87}
\begin{equation}
\label{eq:lpot}
\psi(x,y) = \frac{(b^2)^{1-q}}{2q}(s^2 + (1+\epsilon_c)x^2 + 2\epsilon_s
xy + (1-\epsilon_c)y^2)^q.
\end{equation}
In this expression $b$ is the angular critical radius (\ie the radius
of the critical curve), measured in arc-seconds, for an axisymmetric
($\epsilon_c = \epsilon_s = 0$) lens, $s$ is the angular core radius,
$q$ is a function of the power-law index of the mass density
distribution ($0 \leq q \leq 0.5$), and $\epsilon_c$ and $\epsilon_s$
are related to position angle, $\phi$, of the major axis and the
ellipticity, $\epsilon$, of the potential by:
\begin{align}
\epsilon_c &= \epsilon \cos(2\phi);\\
\epsilon_s &= \epsilon \sin(2\phi).
\end{align} 

It is worth noting that this form for the potential suffers from
several defects.  First, it retains the same ellipticity out to
arbitrary distances; this is in contrast to the expected axisymmetric
behavior at large distances from a mass distribution of finite extent.
A related problem is that the surface mass density distributions
produced by such a potential are unrealistic (``peanut-shaped'') or
unphysical (negative) at large radii.  However, for sufficiently small
$\epsilon$ (\cite{Bla87} suggest $\epsilon < 0.2$) these anomalies
occur at large enough radii that they do not affect the lensing
behavior within the region of interest.  While alternative potentials
based on elliptical mass distributions have been studied, their
lensing behavior is qualitatively the same within the region of
interest as the elliptical potentials \cite[]{Kas93}.  Consequently,
we adopt the simpler potentials for this study.

A potential of the form given in equation~(\ref{eq:lpot}) is
sufficient to describe a system with a single source.  However, for
the case of multiple sources at differing redshifts lensed by the same
mass distribution, $b$ is different for each source.  The constraint
that each source is being lensed by the same mass distribution can be
incorporated into a lensing potential of the form:
\begin{equation}
\label{eq:lpotmulti}
\psi(x,y) = \zeta_k \frac{(b^2)^{1-q}}{2q}(s^2 + (1+\epsilon_c)x^2 + 
2\epsilon_s xy + (1-\epsilon_c)y^2)^q,
\end{equation}
where $b$ is now the same for each source and is equivalent to the
critical radius as $z\rightarrow\infty$.  For the $k$th source
$\zeta_k$ is given by
\begin{equation}
\label{eq:zetak}
\zeta_k = \frac{D_{ls}^{(k)}}{D_s^{(k)}}.
\end{equation}
With this form for the lensing potential the critical radius for each
source is
\begin{equation}
b_k = \zeta_k^{\frac{1}{2-2q}} b.
\end{equation}
Since the angular size distances $D_{ls}^{(k)}$ and $D_s^{(k)}$ are
functions of $\Omega$ and $\lambda$ equation~(\ref{eq:lpotmulti})
allows us to include the cosmological parameters as fitting parameters
in our model, with the distances recalculated each time the
cosmological portion of the model is varied.  

\subsection{Simulation of input data}
\label{sec:simdata}
Our inversion requires two forms of input data: the redshifts of the
lens and arcs and a high-resolution image of the system.  In order to
test the algorithm, simulated data sets were constructed using the
following procedure. First a set of model parameters was selected.
These parameters include the adjustable lens parameters in
equation~(\ref{eq:lpotmulti}), the cosmological parameters, $\Omega$
and $\lambda$, the number of sources to be used, and the redshifts of
the sources.  The lens parameters were chosen to be representative of
clusters of galaxies.  The parameters used in our simulation are
summarized in Table~\ref{tbl:params}. Source positions were
arbitrarily selected but chosen such that strongly lensed arcs are
produced.  Realistic sources were constructed using a representative
galaxy extracted from the Hubble Deep Field \cite[][hereafter
HDF]{Wil96}.  The redshifts for the lens and the four sources were
chosen to be 0.2, 0.4, 0.6, 0.75, and 1.25, respectively.

To incorporate HDF sources, several galaxies from the field were
selected and tested; however, the models produced seemed to differ
little, provided that the sources selected were scaled to a size
appropriate for typical gravitational lensing sources.  Ultimately, a
single source was selected as being representative of the field and
used for all of the sources in our simulations.  Since the redshifts
of most of the HDF sources are unknown, we scaled our source galaxy to
produce simulated arcs of similar width to actual arcs. The background
and peak intensities of these sources were also scaled in order to
produce a desired signal-to-noise ratio in the arcs.

\begin{figure}
\hbox{%
\psfig{figure=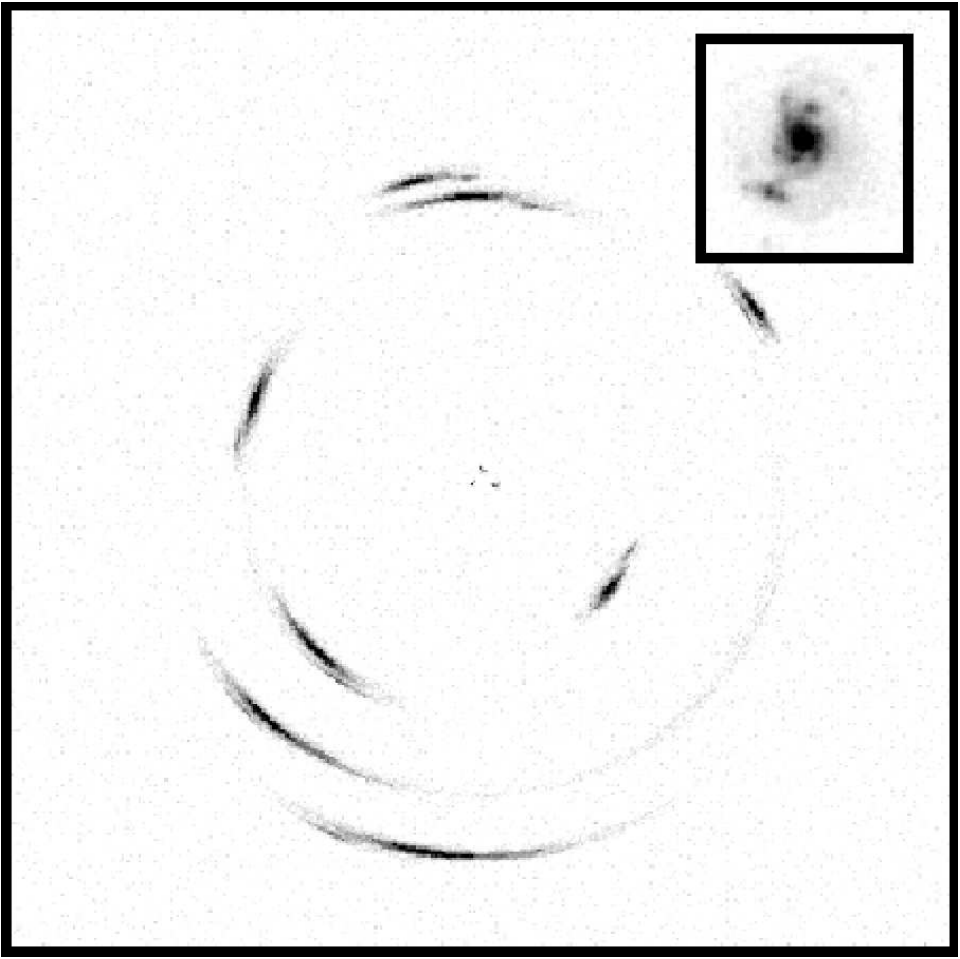,width=5in}}
\caption{The simulated image used in this study.  The inset shows a
source extracted from the Hubble Deep Field and used to construct the
simulated data.  The source was scaled and resampled to produce
sources at each redshift in the simulated system.}
\label{fig:hdfsrc}
\end{figure}

Once model parameters are selected, each pixel in the image plane is
mapped, via the lens mapping equation~(\ref{eq:lensmap}), into each
source plane using the angular-size redshift relation.  Several methods were
considered for selecting the appropriate surface brightness in the
presence of several source planes, including the maximum brightness in
any of the source pixels, the sum of the brightnesses of the source
pixels, and the first (lowest redshift) source pixel encountered with
a brightness above a particular threshold.  In practice, since an
image pixel does not generally map to bright regions in more than one
source plane, there was little difference between the methods.
Therefore, the surface brightness assigned to the image pixel is the
maximum of the surface brightnesses of the source pixels to which it
is mapped in each of the source planes. To produce a desired
signal-to-noise ratio in the image a constant background level is
added to each image pixel, and noise in the form of random Poisson
deviates is added to the simulated image.  The image produced by this
simulation is shown in figure \ref{fig:hdfsrc} (plate 00).

The image produced by the simulation is stored for use as input to our
inversion method.  Also, the bright pixels in the image (\ie the
arcs) are identified with the source plane which produced them (\ie
the process fills the role of measuring the redshifts of real arcs).
These data, along with the redshifts of the source planes, form the
input data for the inversion method.  The model parameters used in the
simulation are stored for comparison to the results of inversion
attempts, but are not otherwise used.

\subsection{Merit function for inversion procedure}
\label{sec:fom}

A numerical model under consideration is evaluated by
computing a merit function, $f$, for the model and comparing it to the
merit functions of other models.  Ideally, the merit function is so
constructed that it has a unique minimum at the ``true'' solution.

\cite{Koc89} proposed a merit function based on the fact that surface
brightness is conserved in gravitational lensing.  We adopt a similar
merit function, but with a few modifications which we find result in
more robust behavior for our multi-source models.  Our merit function
is given by
\begin{equation}
\label{eq:fdefn}
f = \ebar + w_O O + w_S S.
\end{equation}
In equation (\ref{eq:fdefn}) $\ebar$, $O$ and $S$ are functions
defined below which measure the properties that differentiate ``good''
models from ``bad''.  Specifically $\ebar$ measures the degree to
which surface brightness is preserved in the lens mapping; $O$
measures the degree to which the input image data is reproduced by
that reconstructed from the model under consideration; and $S$
measures the degree of fragmentation in the reconstructed sources (\ie
sources should resemble galaxies.)

The quantities $w_O$ and $w_S$ are weights which ensure that the
quantities being summed are all roughly of the same order of magnitude
For the models presented here the weights used were
\begin{align*}
w_O     &= 5,\\
w_S     &= 0.005.
\end{align*}
The `penalty' functions, $O$ and $S$ are designed to enforce physical
constraints upon the numerical minimization algorithm.  In practice
the weight coefficients can be set to zero once the algorithm nears a
solution.

The components of the merit function are calculated as follows; an
image of the system and measurements of the redshifts of the arcs
visible in the system are required as input data.  Arcs are identified
as regions with a surface brightnesses above a particular
``arc-threshold.''  The image plane is divided into regions
corresponding to objects of measured redshift, blank regions, and
regions containing blocking foreground or cluster objects.  At present
we assume that redshifts can be measured for each arc in the system.
This allows each arc to be identified with a particular source plane.
Given a set of model parameters to be evaluated sources are
reconstructed by ray tracing the image pixels back into the source
planes corresponding to the appropriate redshifts.  Each source pixel
is then assigned the average of the values of the image pixels mapped
to that source pixel.  Additionally, multiply imaged pixels in each
source plane are identified.  Background pixels are not used in the
reconstruction of the sources owing to the fact that they have no
measured redshifts, and hence cannot be identified with a particular
source plane.  However, they are used in comparing the input and
reconstructed images as described below.

The function $\ebar$ measures the preservation of surface brightness
and is defined by
\begin{equation}
\label{eq:ebar}
\ebar = \sum_{i,j,k} \frac{1}{\sigma^2_{ij}(n_{ijk}-1)} \sum_{m=1}^{n_{ijk}} 
(I^m_{ijk} - S_{ijk})^2.
\end{equation}
Here, $k$ is summed over the source planes, and $i$ and $j$ are summed
\emph{only} over the multiply imaged pixels in each plane.  $S_{ijk}$
is the value of the source pixel, $n_{ijk}$ is the multiplicity with
which it is imaged, $I^m_{ijk}$ is the value of the $m$th image pixel
identified with the source pixel, and $\sigma_{ij}$ is the noise in
the pixel.

In order to measure how well the image is reproduced, a reconstructed
image is made from the source planes and the model parameters being
evaluated.  The function $O$ is defined by
\begin{equation}
\label{eq:overlap}
O = \frac{1}{N}\sum_{i,j} |I_{ij} - R_{ij}|,
\end{equation}
where $I_{ij}$ refers to the surface brightness of pixel $i,j$ of the
input image, $R_{ij}$ refers to the surface brightness of the same
pixel in the reconstructed image, and $N$ refers to the total number
of pixels in the image.  Since the sources are constructed to
reproduce the observed arcs, the principal contribution to $O$ is from
pixels which were background in the input image, but which are
brighter than the arc-threshold in the reconstruction.  For real data,
regions of the image for which there is no information (\eg a blocking
foreground object or non-transparent portion of the lens) would be
excluded from the calculation of $O$.  The sum is in practice limited
to those pixels whose brightness exceeds the arc-threshold in the
\emph{reconstructed} image in order to prevent Poisson fluctuations in
the background from dominating $O$.

The morphology of galaxies at high redshift is not well known;
however, we expect the surface brightness of neighboring pixels in a
source plane to be correlated.  Thus, we include a contribution to the
merit function which measures the deviation of neighboring pixels in
the source plane.  The function $S$ is defined by
\begin{equation}
S = \sum_{i,j,k} \frac{1}{8} \sum_{m,n=i,j\pm 1} |S_{ijk} - 
S_{mnk}|,
\end{equation}
where the subscripted $S$ refers to pixel values in the various source
planes.  Mathematically, $S$ is the average difference between a pixel
and its eight neighbors in the same source plane.  This statistic
measures the degree to which the source object is fragmented in the
source plane; a contiguous source (\eg one resembling a galaxy) is
preferred over a fragmented source (\eg one resembling a demagnified
version of the arcs in the image).

\subsection{Determination of Model Parameters}
\label{sec:min}
\label{sec:anlytrench}

In order to extract the best fitting model parameters from the system,
we must find the minimum of the merit function over nine parameters.
Unfortunately, we were not able to design an automatic, globally
convergent procedure for finding this minimum.  The principal obstacle
to creating such a procedure is that the merit function surface has
much larger gradients in some principal directions than in others.
Furthermore, far from the solution the merit function surface levels
off to form a plateau with no large-scale gradient to indicate to an
algorithm which way the minimum lies. The first problem can be
overcome by making use of the underlying physical relationships among
the parameters. The second problem can be addressed by noting that for
each of the parameters there is an empirically determined
``characteristic scale'' within which an automatic minimization
algorithm is convergent.  Low-resolution scans can be made of the
parameter space such that suitable initial starting values can be
chosen.  The characteristic scales for the model parameters were
determined from an extensive search of the parameter space and are
given in Table~\ref{tbl:params}.

\subsubsection{Computational Techniques}

To fully exploit the information contained within HST images of arc
systems requires the use of a high-resolution grid of image and source
pixels.  The many evaluations of the merit function in a minimization
algorithm places large demands on computer resources.  The work
presented here was performed using the SCAAMP\footnote{Scientific
Applications on Arrays of Multiprocessors} collaboration's
64-processor Silicon Graphics Origin 2000.  The facility has maximum
throughput of 21 Gflops, although our calculations typically used only
16 of the 64 processors. The availability of this resource allowed us
to thoroughly study the parameter space in a reasonable amount of time.
Details regarding the software developed for this procedure will be
presented at a later date \cite[]{link}.

\begin{figure}
\hbox{%
\psfig{figure=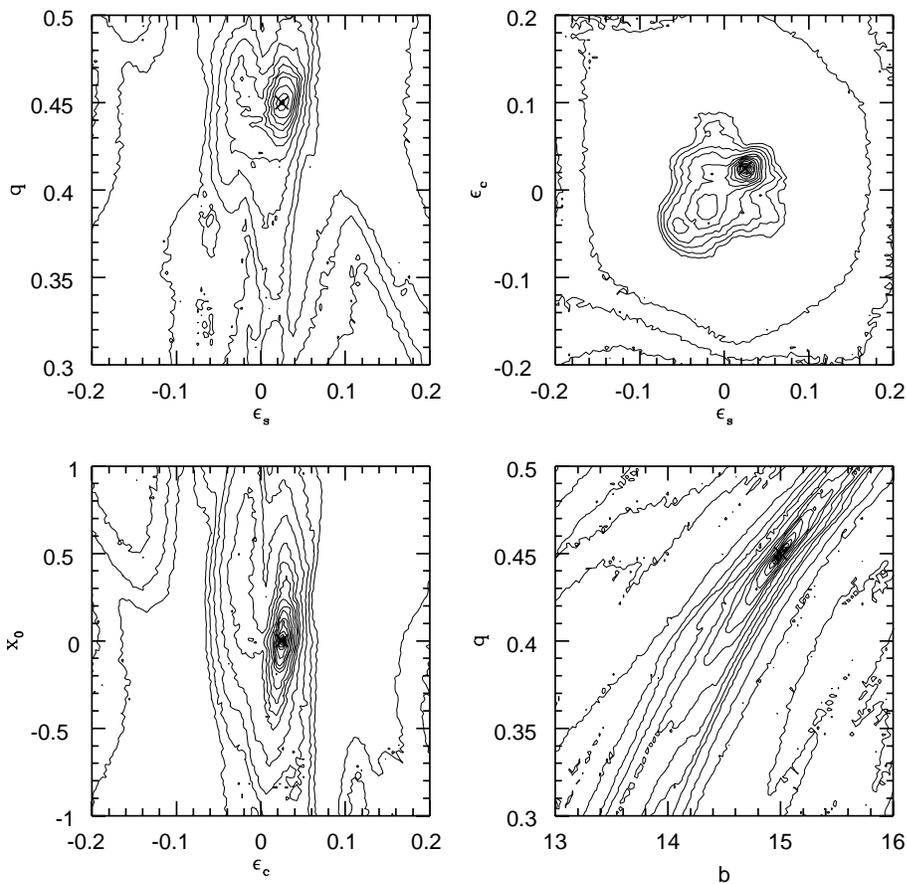,width=5in}}
\caption{Contours of the merit function, $f$, for a selection of 2-D
slices in the parameter space.  In each case the cross marks the
fiducial parameters used in the simulation.  Note the correlation
between the power-law index of the lens potential, $q$, and the
overall scale factor, $b$ (\qv equation~\protect\ref{eq:lpotmulti}).}
\label{fig:slices}
\end{figure}

A high-resolution scan of the parameter space was performed in order
to determine the characteristic scales of the model parameters
(Table~\ref{tbl:params}).  We believe these scales will be generally
applicable to gravitational lens systems parameterized in the form of
equation~(\ref{eq:lpotmulti}).  Since the characteristic scales
determine the size of the neighborhood of convergence, a scan of
parameter space at this resolution can be used to select initial
guesses for a minimization algorithm.  Some representative slices of
the parameter space are shown in figure~\ref{fig:slices}.

Having selected initial guesses we fix the cosmological parameters at
reasonable values, the remaining
parameters are determined using a standard downhill simplex
technique \cite[]{Pre88}.   Since the cosmological parameters
have a small, second-order effect on the model we find that
it is possible to arrive at a good solution for the lens parameters
regardless of the initial guesses for cosmological parameters.

Once  the mass parameters for the system are obtained we solve for the
cosmological parameters.  Although small, the cosmological effects are
strongly coupled only to the effects of the $b$ parameter (\cf
equation~\ref{eq:lpotmulti}.)  Thus, with the exception of $b$, the
lens mass parameters are held constant while the cosmological
parameters are being varied.

The correlation between $b$ and the cosmological parameters can be
understood as a tradeoff between the terms multiplying the front of
the right side of equation~(\ref{eq:lpotmulti}).  This relationship
can be used to assist the minimization algorithm in finding the cosmological
solution once the mass parameters have been found.  To further
understand this relation, consider a simple system with only one
source.  Let us, however, describe this system using the form of the
lensing potential given in equation (\ref{eq:lpotmulti}).  Then, all
other things being equal, models for which
\begin{equation}
\frac{D_{ls}}{D_s} b^{2(1-q)} = \text{constant}
\end{equation}
will produce identical arc systems.  In the case of multiple source
models we find that the minimal surface can be approximated by
\begin{equation}
\label{eq:anly}
\zeta b^{2(1-q)} = \Upsilon
\end{equation}
where $\Upsilon$ is a constant, and $\zeta$ represents the distance
ratio, $D_{ls}/D_{s}$ for some sort of ``characteristic'' redshift of
the sources.  We find that by taking this characteristic redshift to
be the median redshift of all of the sources, a reasonable
approximation of the relationship between $b$ and $\Omega$ (with other
parameters constant) can be obtained in the neighborhood of the
solution.  However, since this approximation is strictly true only for
the median redshift, we use equation~(\ref{eq:anly}) only to provide
an initial guess for a one-dimensional minimization in $b$ at each
evaluation of the merit function.

\section{Results}
\label{sec:rslts}

We find that the minimization procedure outlined in the previous
section converges to the global minimum of the merit function if the
initial guesses for the model parameters are within a ``neighborhood
of convergence'' about the correct values. This neighborhood is
readily identifiable from the low-resolution parameter space scans
described in section~\ref{sec:min}.  For a signal-to-noise ratio of 10
in the simulated data the lens parameters recovered by the algorithm
are essentially indistinguishable from their fiducial values (\qv
Table~\ref{tbl:params}), while the cosmological parameters recovered
are accurate to within $\sim$10\%.  

\begin{figure}
\hbox{%
\psfig{figure=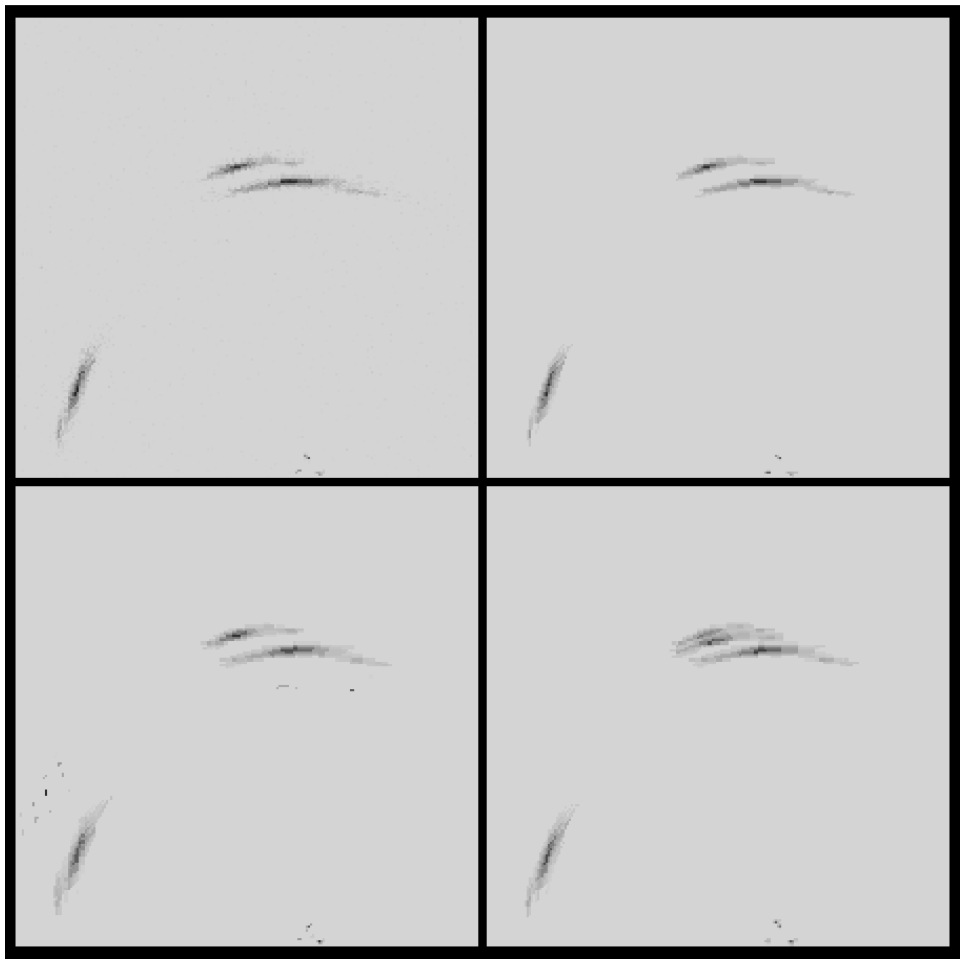,width=5in}}
\caption{A comparison of a fixed region within (clockwise from
upper-left) the simulated image, the image reconstructed from the best
fit model (see also Table~\protect\ref{tbl:params}), and the images
reconstructed from the best fit models with cosmological parameters
fixed at $\Omega=0$, $\lambda=1$, and $\Omega=0.2$, $\lambda=0$,
respectively.  Note that the best fit model reproduces the simulated
image well, while the others have noticeable defects.}
\label{fig:imgmosaic}
\end{figure}

\begin{figure}
\hbox{%
\psfig{figure=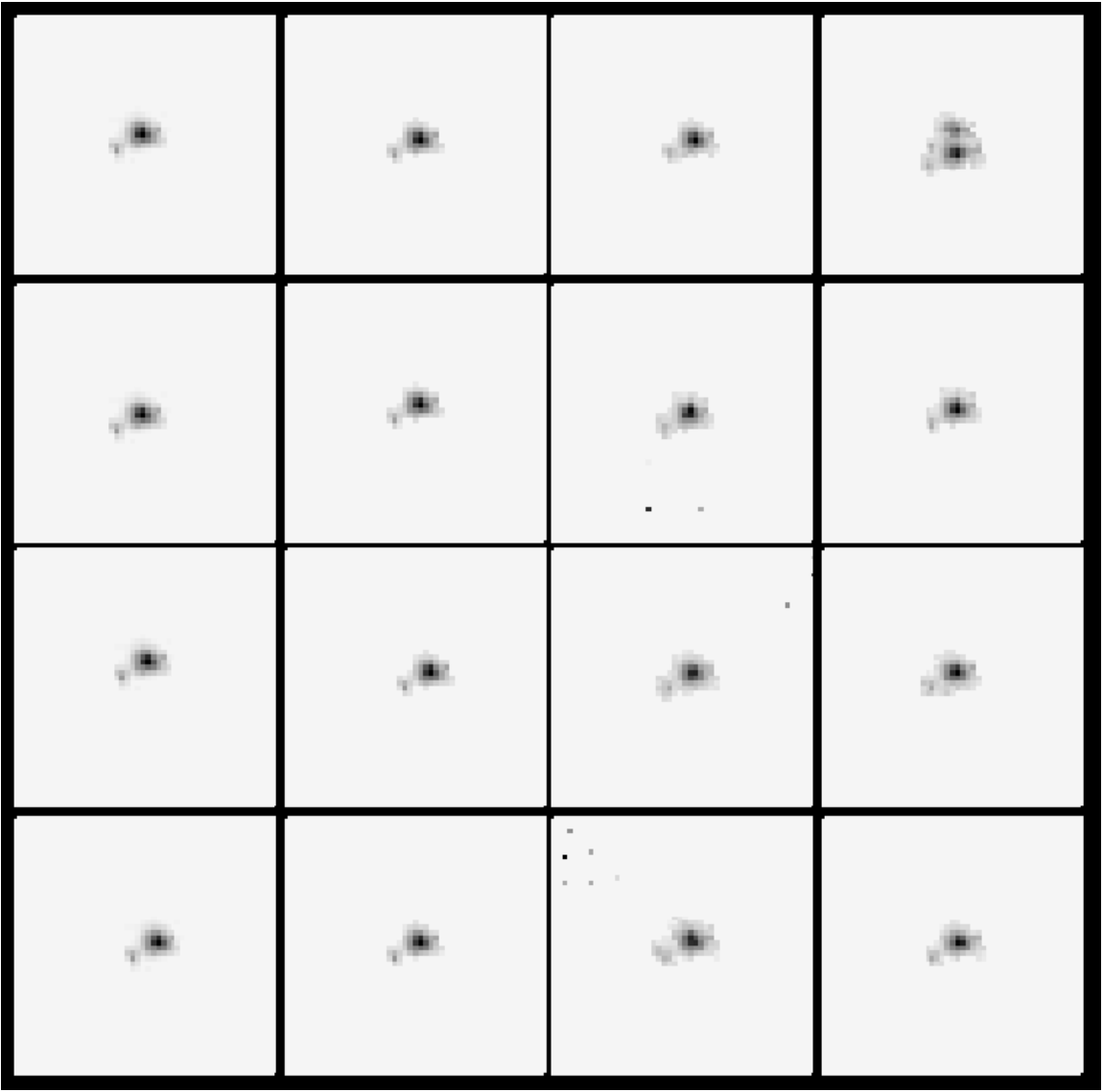,width=6in}}
\caption{A comparison of the input data and reconstructed source
planes for the models shown in figure~\protect\ref{fig:imgmosaic}.
From top to bottom are the four source planes corresponding to the
redshifts of the arcs in the system.  From left to right are the input
data, the best fit model, and the best fit models with $\Omega=0$,
$\lambda=1$ and $\Omega=0.2$, $\lambda=0$.  Note that as was the case
for the reconstructed images, the best fit model reproduces the
simulated data well, while there are noticeable defects in the sources
corresponding to the other models, particularly at the two redshift
extremes.}
\label{fig:srcmosaic}
\end{figure}

We can gain some insight into the cosmological sensitivity of our
technique by examining the morphology of sources and images
reconstructed from a selection of cosmological models.  In
figures~\ref{fig:imgmosaic} and \ref{fig:srcmosaic} (plates 00,00) we
compare the simulated data with that reconstructed from both the best
fit model and models with the cosmological parameters fixed at $\Omega
= 0$, $\lambda=1$ and $\Omega = 0.2$, $\lambda=0$.  In each of the
latter two cases the non-cosmological parameters were varied in order
to produce the best fit model for the cosmological model under
consideration.  Figure~\ref{fig:imgmosaic} shows a portion of the
simulated image and corresponding regions of the image reconstructed
from the model recovered by the algorithm and of images reconstructed
from models with the cosmological parameters fixed at $\Omega=0$,
$\lambda=1$ and $\Omega=0.2$, $\lambda=0$.  Figure~\ref{fig:srcmosaic}
shows the four source planes corresponding to each of these images.
While the images and sources for the incorrect cosmological models are
similar to the originals, they contain noticeable defects.  In
contrast, the reconstructions from the model arrived at by the
minimization algorithm appear substantially better.

\begin{figure}
\hbox{%
\psfig{figure=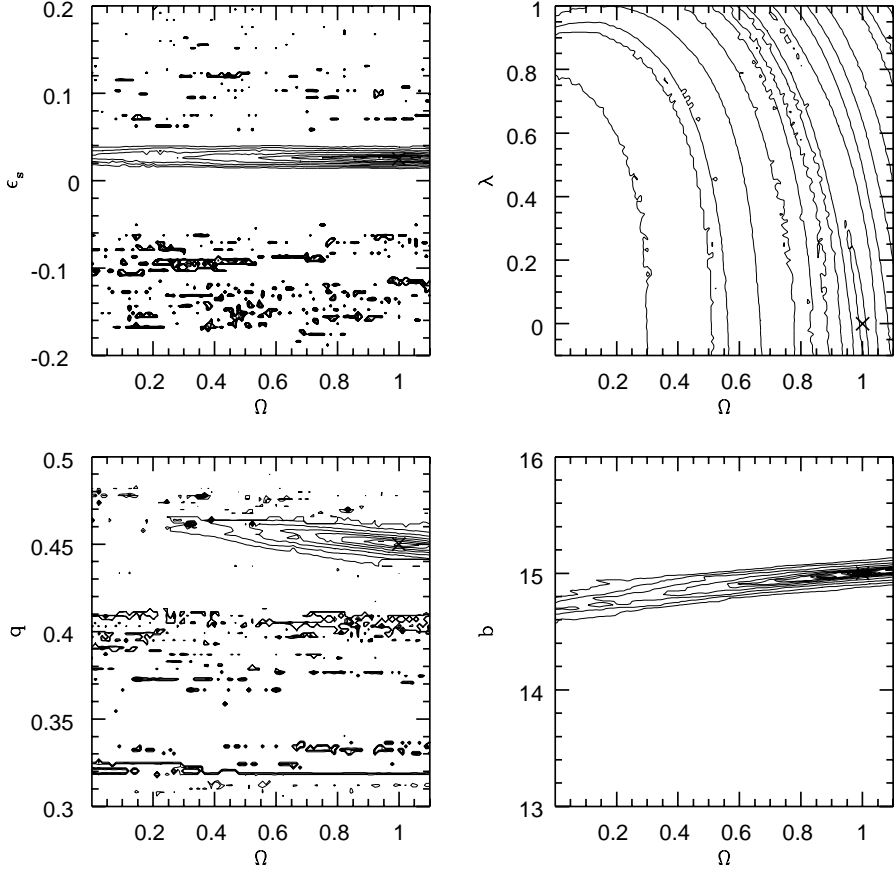,width=5in}}
\caption{Contours of the merit function, $f$, for parameter space
slices including the cosmological parameters ($\Omega$ and $\lambda$).
In slices not involving $b$, we adopt the value of $b$ which minimizes
$f$ with other parameters fixed.  In each case the cross marks the
parameters used in the simulation.  Note in panel (d) the correlation
between $b$ and $\Omega$ which results from the fact that both
parameters affect the overall scale of the system.  Note also the
existence of a well-defined minimum in the cosmological parameters.
This demonstrates the ability of this technique to discriminate
between cosmological models.}
\label{fig:cosmslices}
\end{figure}

Figure~\ref{fig:cosmslices} shows contours of the merit function
involving the cosmological parameters.  Note that the cosmological
parameters show strong correlations with $b$.  This is due to the fact
that these parameters affect the overall scale of the system through
the $\zeta_k$ in equation~(\ref{eq:lpotmulti}).  Hence, for the first
three panels $b$ was adjusted to produce the minimal value of $f$ at
each point in the contour.  This allows us to isolate the intrinsic
effects of the cosmological parameters from those that result from
changing the overall multiplicative constant in the lensing potential.

\begin{figure}
\hbox{%
\psfig{figure=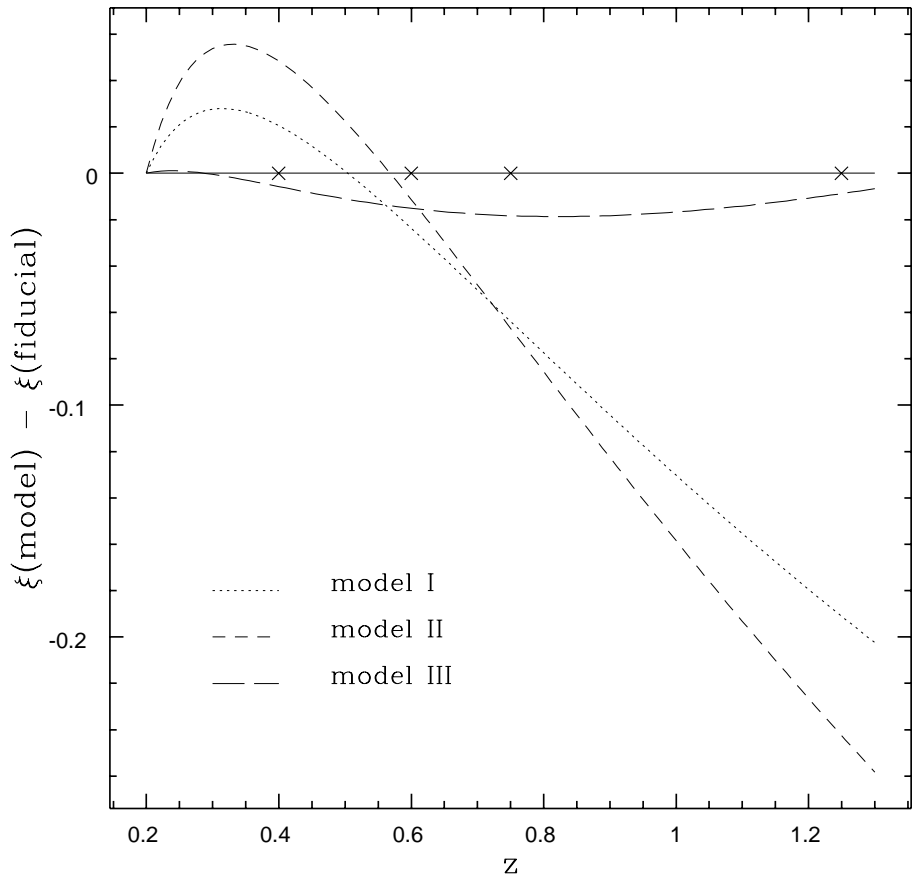,width=5in}}
\caption{Residuals of the quantity $\xi$ in
equation~\protect\ref{eq:interp} with respect to the fiducial model as
a function of redshift.  The crosses mark the redshifts of the arcs in
our simulated data.  Models which produce small residuals at these
redshifts should produce ``better'' fits with this technique. Model I
is a flat, low-omega model, model II is a $\lambda=0$, low-omega
model, and Model III is a model along the curve of the elongated
contours of $f$ in the $\Omega$--$\lambda$ plane
(figures~\protect\ref{fig:cosmslices} and \protect\ref{fig:monte}).
Note that model III, which produces the best fit of these three, also
produces the smallest residuals of $\xi$ with respect to the fiducial
model.}
\label{fig:interp}
\end{figure}

Of particular interest is the slice in $\Omega$ and $\lambda$, since
this characterizes our cosmological sensitivity.  For our particular
simulation the cosmological sensitivity is primarily to $\Omega$
rather than $\lambda$.  This is partly a result of the fiducial model
we have chosen for this study and partly a consequence of the range of
redshifts chosen for the simulation.  By choosing $\Omega = 1$,
$\lambda = 0$, a strongly mass-dominated model, for our fiducial
cosmology, we effectively mask the influence of $\lambda$ unless
$\lambda$ is comparable to $\Omega$.  Moreover, the angular
size-redshift relation, upon which this technique depends for its
cosmological sensitivity, is sampled only over the range redshifts of
spanned by lens and the sources.  Consequently, cosmological models
which produce similar angular size-redshift relations \emph{over those
redshifts} will ``look'' similar.
Examination of equation~(\ref{eq:lpotmulti}) reveals that the
cosmological parameters enter into the lensing potential through the
leading multiplicative term:
\begin{equation}
\label{eq:interp}
\xi = \frac{b^{2(1-q)}}{2q} \zeta(z)
\end{equation}
Figure~\ref{fig:interp} illustrates the dependence of this expression
on redshift for selected cosmological models.  For each cosmological
model we evaluate the quantity $\xi$ in equation~(\ref{eq:interp}),
using the values of $b$ and $q$ which minimize $f$, and compare this
to the same quantity evaluated for the fiducial cosmology.  The
differences are plotted as a function of redshift for each
cosmological model.  For comparison we show three cosmological models
of interest: model I, a flat, low-omega model; model II, a
$\lambda=0$, low-omega model; and model III, a model along the
near-minimal surface of $f$ in the $\Omega$-$\lambda$ plane.
Note that the residuals for model III are substantially smaller than
those for the other models over the range of redshifts represented in
the simulated data.  Thus, it is understandable that model III is a
better fit (as measured by its relatively low value of $f$) than
models I and II.  Therefore, the shape of the merit function contours
in the $\Omega$--$\lambda$ plane results from the behavior of
equation~(\ref{eq:interp}) over the range of redshifts sampled by the
lens and sources.  Specifically, the contours are elongated in the
direction of cosmological models that produce values of $\xi$ similar
to those of the fiducial model at the sampled redshifts.

\subsection{Monte carlo estimate of confidence regions}
\label{sec:monte}

We have identified four principal sources of error in the parameters
determined using our method.  These are: the noise in the input image,
errors in the redshift measurements, deviation of the actual lens
potential from our parameterized form, and any density fluctuations
along the line of sight which alter the angular size-redshift
relation. We consider only the first two sources of error in this
study; although the others will be addressed in subsequent papers in
this series.

\begin{figure}
\hbox{%
\psfig{figure=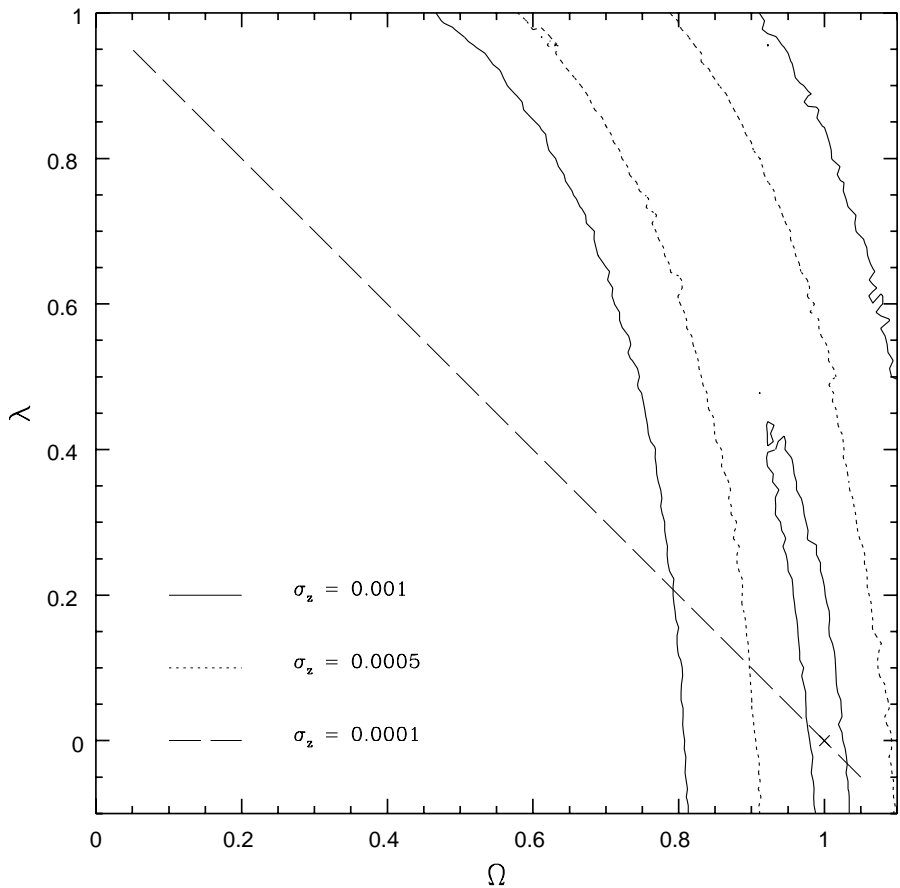,width=5in}}
\caption{A contour of the merit function, $f$, in the
$\Omega$--$\lambda$ subspace showing the confidence region estimated
for several values of $\sigma_z$ using the procedure in
section~\ref{sec:monte}.  For $\sigma_z=0$ (\ie no measurement errors
in the redshifts of the arcs) the estimated confidence region is
approximately the size of the cross which marks the fiducial model.
Although the allowed region is fairly large with typical redshift
errors, along the directions of astronomical interest
($\Omega+\lambda=1$, shown by the dashed line, and $\lambda = 0$) the
constraints are quite strong (\qv Table~\protect\ref{tbl:monte}).}
\label{fig:monte}
\end{figure}

In order to evaluate these first two sources of error, we have
performed monte carlo calculations to estimate confidence regions in
the cosmological parameters.  At each iteration of these calculations
new simulated data are constructed from the fiducial model described
in the previous sections, using a new seed to create an independent
image of the system with the same signal-to-noise ratio as the
original.  Next, errors with a Gaussian distribution are added to the
redshifts in the model, and the merit function, $f$, is calculated for
the fiducial parameters and the perturbed redshifts.  The process is
repeated until a sufficiently large sample is collected; in this case
100 samples were used.  The contour in the $\Omega$--$\lambda$
parameter space corresponding to the 68.5 percentile of $f$ is adopted
as an estimate for the $1\sigma$ confidence region for $\Omega$ and
$\lambda$.  Figure~\ref{fig:monte} shows the $1\sigma$ confidence
regions obtained through this technique for several values of
$\sigma_z$, the standard error in the redshifts.  With a peak
signal-to-noise ratio of ten in the image arcs, the contribution from
the measurement errors in the redshifts is by far the dominant effect.

If we adopt $\sigma_z = 0.001$ as a typical error in current
measurements the allowed region of the $\Omega$--$\lambda$ parameter
extends over a broad range in $\lambda$ and a modest range in
$\Omega$.  However, much of the parameter space is nevertheless
excluded.  In particular, our method strongly strongly discriminates
between ``interesting'' cosmological models, viz. flat cosmologies
($\Omega+\lambda=1$) and $\lambda=0$ cosmologies.  For these classes
of models the constraints from this approach are strong.  Results for
these specific cases are summarized in Table~\ref{tbl:monte}.  It is
worth noting that the allowed range of cosmological parameters is
significantly reduced if the error in the arc redshifts can be reduced
to $\sigma_z \approx 0.005$.  While such measurements are difficult,
they are not beyond the reach of 10-m class telescopes.

\section{Concluding Remarks}

Our inversion technique is able to extract both the lens and the
cosmological parameters used to construct the simulated data.
Moreover, the confidence regions obtained in our monte carlo
calculation suggest that physically interesting constraints could be
set on cosmological parameters through this method.  The requisite
data, high-resolution images from HST and redshifts for the arcs, are
are currently obtainable. We are encouraged by recent observations
which include systems with multiple sources \cite[]{Lup94,Kne96}, and
by the fact that some systems appear to have sources at several
redshifts \cite[]{Hog96}.

%xxx Change 5
We have made several simplifying assumptions in our analysis,
particularly the use of the same analytic potential for both
simulating and analyzing the data.  We are currently engaged in
further work studying the cosmological constraints which result from
applying this technique to a potential whose form is not known \emph{a
priori}, and which may contain substructure.  Presumably these factors
will introduce additional error which may weaken the constraints which
can be set on cosmological parameters using this technique.  However,
in view of the success we have experienced with our simulations to
date we believe that the errors so introduced will be surmountable and
that this method is capable of placing interesting bounds on
cosmological parameters.  

\section*{Acknowledgements}

The SCAAMP collaboration is supported by the Office of
Cross-Disciplinary Activities of the CISE directorate of the National
Science Foundation through the Academic Research and Infrastructure
Grant CDA-9601632.

%\bibliography{mnemonic,gl}

\clearpage

\begin{thebibliography}{}

\bibitem[Blandford \& Kochanek(1987)Blandford and Kochanek]{Bla87}
Blandford, R.~D., \& Kochanek, C.~S. 1987, ApJ, 321, 658

\bibitem[Blandford \& Narayan(1992)Blandford and Narayan]{Bla92}
Blandford, R.~D., \& Narayan, R. 1992, ARAA, 30, 311

\bibitem[Breimer \& Sanders(1992)Breimer and Sanders]{Bre92}
Breimer, T.~G., \& Sanders, R.~H. 1992, MNRAS, 257, 97

\bibitem[Cen {et~al.}(1994)Cen, Gott, Ostriker, and Turner]{Cen94}
Cen, R., Gott, J.~R., Ostriker, J.~P., \& Turner, E.~L. 1994, ApJ, 423, 1

\bibitem[{Hogg} {et~al.}(1996){Hogg}, {Blandford}, {Kundic}, {Fassnacht}, and
  {Malhotra}]{Hog96}
{Hogg}, D.~W., {Blandford}, R., {Kundic}, T., {Fassnacht}, C.~D., \&
  {Malhotra}, S. 1996, \apjl, 467, L73

\bibitem[Kassiola \& Kovner(1993)Kassiola and Kovner]{Kas93}
Kassiola, A., \& Kovner, I. 1993, ApJ, 417, 450

\bibitem[Kneib {et~al.}(1996)Kneib, Ellis, Smail, Couch, and Sharples]{Kne96}
Kneib, J.~P., Ellis, R.~S., Smail, I., Couch, W.~J., \& Sharples, R.~M. 1996,
  ApJ, 471, 643

\bibitem[Kochanek {et~al.}(1989)Kochanek, Blandford, Lawrence, and
  Narayan]{Koc89}
Kochanek, C.~S., Blandford, R.~D., Lawrence, C.~R., \& Narayan, R. 1989, MNRAS,
  238, 43

\bibitem[Link(1998)Link]{link}
Link, R. 1998.
\newblock Ph.D. thesis, Indiana University

\bibitem[Luppino {et~al.}(1994)Luppino, Gioia, Annis, Le~F\`evre, and
  Hammer]{Lup94}
Luppino, G.~A., Gioia, I.~M., Annis, J., Le~F\`evre, O., \& Hammer, F. 1994,
  ApJ, 416, 444

\bibitem[Lynds \& Petrosian(1989)Lynds and Petrosian]{Lyn88}
Lynds, R., \& Petrosian, V. 1989, ApJ, 336, L1

\bibitem[Paczynski(1987)Paczynski]{Pac87}
Paczynski, B. 1987, Nature, 325, 572

\bibitem[Press {et~al.}(1988)Press, Teukolsky, Vetterling, and Flannery]{Pre88}
Press, W.~H., Teukolsky, S.~A., Vetterling, W.~T., \& Flannery, B.~P. 1988.
\newblock Numerical Recipes: the Art of Scientific Computing, Cambridge
  University Press

\bibitem[Sandage(1988)Sandage]{San88}
Sandage, A. 1988, ARAA, 26, 561

\bibitem[Schneider, Ehlers, \& Falco(1992)Schneider, Ehlers, and Falco]{Sch92}
Schneider, P., Ehlers, J., \& Falco, E.~E. 1992.
\newblock Gravitational Lenses, Springer-Verlag

\bibitem[Soucail {et~al.}(1987)Soucail {\em et~al.}]{Sou87}
Soucail, G., et~al. 1987, A. Ap., 184, L7

\bibitem[Turner(1990)Turner]{Tur90}
Turner, E.~L. 1990, ApJ, 365, L1

\bibitem[Williams {et~al.}(1996)Williams {\em et~al.}]{Wil96}
Williams, R.~E., et~al. 1996, \aj, 112

\end{thebibliography}

\clearpage

\clearpage

\begin{deluxetable}{rll}
\tablewidth{344pt}
\tablecaption{Simulation Parameters\label{tbl:params}}
\tablehead{
\colhead{Parameter} &
\colhead{Characteristic Length} &
\colhead{Fiducial Value\tablenotemark{a}}
}
\startdata
$\Omega$	& large\tablenotemark{b}         & 1.0 	 \nl
$\lambda$	& large\tablenotemark{b}         & 0.0 	 \nl 
$b$		& 1\farcs 5       & 15\farcs 0 	  \nl
$q$		& 0.05            & 0.45 	        \nl
$s$		& 1\arcsec        & 2\farcs 5 	   \nl
$\epsilon_c$	& 0.05            & 0.025 	         \nl
$\epsilon_s$	& 0.05		  & 0.025 	         \nl
$x_0$		& 1\arcsec        & 0\farcs 0 	   \nl
$y_0$		& 1\arcsec        & 0\farcs 0 	   \nl
\enddata
\nl

\tablenotetext{a}{The values of the parameters recovered by this
method (see text) are indistinguishable from their fiducial values.}

\tablenotetext{b}{``Large'' means that the characteristic length of the
parameter extends to the entire physical range of the parameter.}

\end{deluxetable}

\clearpage

\begin{deluxetable}{rrr}
\tablewidth{0pt}
\tablecaption{Cosmological Bounds vs. Redshift Errors.\label{tbl:monte}}
\tablehead{
\colhead{Case} &
\colhead{$\sigma_z = 0.001$} &
\colhead{$\sigma_z = 0.0005$}
}
\startdata
$\Omega + \lambda = 1$ & $\Omega > 0.8$, $\lambda < 0.2 $ & $\Omega >
0.9$, $\lambda< 0.1$ \nl
$\lambda = 0$	& $0.8 < \Omega < 1.2$	& $0.9 < \Omega < 1.1$ \nl
$\Omega = 1$	& $\lambda < 0.8$	& $\lambda < 0.5$ \nl
\enddata
\nl
\end{deluxetable}

\end{document}